\documentclass{article}

\usepackage{natbib}
\usepackage{microtype}
\usepackage{graphicx}
\usepackage{booktabs} 
\usepackage{xspace}
\usepackage{enumitem}
\usepackage{csvsimple}
\usepackage[skip=0pt,labelfont=bf]{caption}
\usepackage{subcaption}
\usepackage{amsmath}
\usepackage{tcolorbox, xcolor, soul, xspace}
\usepackage[cachedir=minted-cache]{minted}
\usepackage{amssymb}

\setlist[itemize]{leftmargin=*, noitemsep, topsep=0pt}

\usepackage{hyperref}

\usepackage{tcolorbox}
\tcbuselibrary{listings, breakable, skins, theorems}
\usepackage{soul} 

\sethlcolor{yellow!30}

\usepackage{booktabs}
\usepackage{colortbl}
\usepackage{xcolor}

\rowcolors{2}{gray!5}{white}

\usepackage{pgfplotstable}
\usepgfplotslibrary{colorbrewer}
\pgfplotsset{compat=1.17}

\definecolor{OliveGreen}{rgb}{0,0.6,0}

\newcommand{\noindentbf}[1]{\noindent \textbf{#1}\xspace}
\newcommand{\benchName}{\textsc{gpuFLOPBench}\xspace}
\newcommand{\globalKernel}{\texttt{\_\_global\_\_}\xspace}
\newcommand{\deviceKernel}{\texttt{\_\_device\_\_}\xspace}

\newcommand{\gptfouromini}{\texttt{gpt-4o-mini}\xspace}
\newcommand{\othreemini}{\texttt{o3-mini}\xspace}
\newcommand{\gptfivemini}{\texttt{gpt-5-mini}\xspace}

\newcommand{\easySubset}{\textit{easy}\xspace}
\newcommand{\hardSubset}{\textit{hard}\xspace}

\newcommand{\noFLOPs}{\textsc{No FLOPs}\xspace}
\newcommand{\spOnly}{\textsc{SP-Only}\xspace}
\newcommand{\dpOnly}{\textsc{DP-Only}\xspace}
\newcommand{\mixed}{\textsc{Mixed}\xspace}

\usepackage[accepted,nohyperref]{mlsys2025}

\mlsystitlerunning{Counting Without Running: Evaluating LLMs’ Reasoning About Code Complexity}

\begin{document}

\twocolumn[
\mlsystitle{Counting Without Running: Evaluating LLMs’ Reasoning About Code Complexity}



\mlsyssetsymbol{equal}{*}

\begin{mlsysauthorlist}
\mlsysauthor{Gregory Bolet}{vt}
\mlsysauthor{Giorgis Georgakoudis}{llnl}
\mlsysauthor{Konstantinos Parasyris}{llnl}
\mlsysauthor{Harshitha Menon}{llnl}
\mlsysauthor{Niranjan Hasabnis}{codemetal}
\mlsysauthor{Kirk W. Cameron}{vt}
\mlsysauthor{Gal Oren}{stanford,technion}
\end{mlsysauthorlist}

\mlsysaffiliation{vt}{Virginia Tech, Blacksburg, Virginia, USA}
\mlsysaffiliation{llnl}{Lawrence Livermore National Laboratory, Livermore, California, USA}
\mlsysaffiliation{codemetal}{Codemetal, Boston, Massachusetts, USA}
\mlsysaffiliation{stanford}{Stanford University, Stanford, California, USA}
\mlsysaffiliation{technion}{Technion, Haifa, Israel}

\mlsyscorrespondingauthor{Gregory Bolet}{gbolet@vt.edu}

\mlsyskeywords{Machine Learning, MLSys}

\vskip 0.3in

\begin{abstract}
Modern GPU software stacks demand developers who can anticipate performance bottlenecks before ever launching a kernel; misjudging floating-point workloads upstream can derail tuning, scheduling, and even hardware procurement.
Yet despite rapid progress in code generation, today’s Large Language Models (LLMs) are rarely tested on this kind of forward-looking reasoning.
We close that gap with \benchName, a benchmark that asks models to ``count without running'' by predicting single and double-precision FLOP counts for 577 CUDA kernels drawn from HeCBench, annotated with ground-truth profiles and eight execution attributes that distinguish trivially analyzable code from kernels whose FLOPs depend on hidden compiler or runtime behavior.
Evaluating current closed-source reasoning models shows clear but uneven progress: the newest LLMs achieve perfect classification on straightforward kernels but still incur multiple order-of-magnitude errors whenever implicit FLOPs arise from division, intrinsic math functions, or common subexpressions.
These results surface a core limitation of existing code assistants -- the inability to internalize hardware-specific microcode effects -- and position \benchName as a focused testbed for developing LLM tooling that can reason about performance with the same rigor as experienced GPU developers.  Sources are available at our repository:
\url{https://github.com/Scientific-Computing-Lab/gpuFLOPBench}.
\end{abstract}

]



\printAffiliationsAndNotice{\mlsysEqualContribution} 

\section{Introduction}

Modern performance-critical software stacks depend on developers being able to anticipate the computational load of a kernel before it ever runs.
The HPC community routinely relies on FLOP budgets to choose algorithms, schedule kernels on constrained accelerators, and reason about roofline trade-offs \cite{chenLandscapeChallengesHPC2024,rooflineModel}.
Yet despite rapid progress in LLM-based code generation \cite{chen2021evaluating,austin2021program}, we lack evidence that these models internalize the analytic reasoning experts use to judge code efficiency.
We therefore ask: \textit{can present-day LLMs estimate the floating-point work performed by a GPU kernel purely from its source?}

Existing evaluations provide only partial signals.
General-purpose benchmarks such as HumanEval, MBPP, and LiveCodeBench verify functional correctness but rarely penalize inefficient solutions \cite{chen2021evaluating,austin2021program,jainLiveCodeBenchHolisticContamination2024}.
Competitive-programming datasets like APPS and CodeContests demand performance indirectly through runtime limits \cite{hendrycks2021apps,li2022alphacode}.
Specialized GPU efforts -- including \textsc{KernelBench}, \textsc{MultiKernelBench}, and \textsc{TritonBench} -- measure realized speedups after execution rather than asking whether a model can anticipate cost beforehand \cite{ouyang2025kernelbench,wen2025multikernelbench,li2025tritonbench}.
We delve into these studies in \autoref{related_work}; here, we highlight the missing capability: static reasoning about performance.

\begin{figure*}[ht!]
    \centering
    \includegraphics[width=0.95\linewidth,trim={0 0cm 0 0},clip]{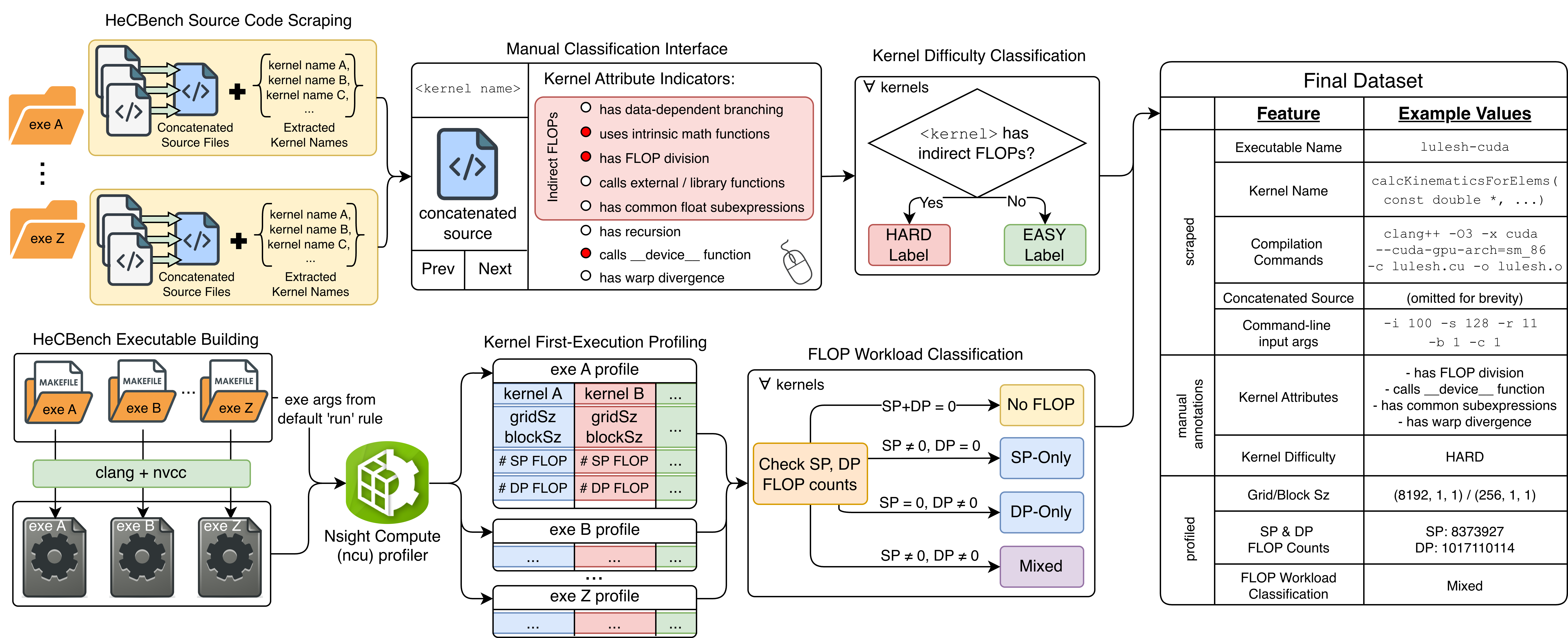}
    \caption{Overview of dataset creation steps \benchName. 
    We scrape CUDA kernels from the HeCBench suite and profile the first invocations of the kernels, gathering: kernel names, source codes, grid and block size, kernel launch params, command-line execution arguments, and profiled Single/Double-precision FLOP counts.
    Each kernel's source code is manually classified as having execution attributes that could introduce implicit/indirect FLOPs, leading to its categorization as a \textit{hard} code that cannot easily be directly statically analyzed.
    Example values from the \texttt{lulesh-cuda} program are shown for clarification.
    }
    \label{fig:datasetCreation}
\end{figure*}

We address this gap with \benchName, a benchmark that asks models to ``count without running.''
For each of 577 CUDA kernels drawn from HeCBench \cite{hecbenchPaper,hecbenchGithub}, we provide the kernel source, launch configuration, and command-line arguments, and pair them with ground-truth single- and double-precision FLOP counts captured from first invocations on NVIDIA hardware.
We manually annotate eight execution attributes that capture when FLOPs depend on compiler or runtime effects, partitioning the dataset into \textit{easy} and \textit{hard} subsets.
\autoref{fig:datasetCreation} outlines the profiling and annotation pipeline.

We evaluate a range of contemporary closed-source reasoning models on this task, querying each model with a structured prompt that requires both counts and explanations.
The newest systems (e.g., GPT-5-mini) perfectly classify straightforward kernels, yet all models exhibit order-of-magnitude Mean Absolute Log Error once implicit FLOPs from division, template instantiation, or library calls appear.
These behaviors indicate that, while progress is tangible, current LLMs still lack an internal cost model that accounts for hardware-specific microcode effects.

\textbf{Contributions.} In summary, this paper makes the following contributions:

\textbf{Benchmark Dataset.} 
We curate \benchName, the first dataset of real-world CUDA kernels with paired single- and double-precision FLOP counts and execution-attribute annotations, enabling controlled tests of static performance reasoning.

\textbf{Evaluation Framework.}
We design a prompting and analysis pipeline that elicits structured FLOP predictions from LLMs and measures both categorical workload classification and regression error (MALE) on SP/DP counts.

\textbf{Empirical Findings.}
We benchmark several proprietary reasoning models and document where they succeed (regular control flow) and fail (implicit FLOPs), revealing systematic blind spots traceable to vendor-specific execution semantics.

\textbf{Broader Implications.}
By isolating static FLOP prediction, \benchName offers a focused lens on an emerging requirement for coding assistants: understanding why code performs the way it does, not just producing syntax-correct solutions.

Ultimately, our work highlights a new frontier for evaluating and improving coding LLMs --- the shift from ``Can it write code that works?'' to ``Does it understand why the code works and how efficient it is?''
Bridging this understanding is essential for the next generation of AI coding assistants that can debug, optimize, and explain code at an expert level.

\section{Related Work}
\label{related_work}

We discuss recent work related to LLMs and performance prediction along three different dimensions: (1) evaluation of LLMs' ability to predict program performance, (2) ML-based performance prediction, and (3) existing benchmarks for performance prediction.

\subsection{LLMs for Code Performance Prediction}

Recent studies have begun probing whether LLMs understand code performance beyond pass/fail correctness.
\textsc{AlphaCode}~\cite{li2022alphacode} showed that large transformers can solve competitive programming problems at mid--pack human levels, but relied on massive sampling and filtering rather than intrinsic reasoning about efficiency.
Inefficient solutions were eliminated through test--case failures, not analytic understanding.
In contrast, this work directly tests an LLM’s ability to reason about algorithmic complexity without execution feedback.

Static--reasoning benchmarks are emerging to assess such capabilities.
Xie et~al.\ introduce \textsc{CoRe}~\cite{xie2025core}, evaluating LLMs on static--analysis tasks like identifying data and control dependencies.
While models detect shallow patterns, they struggle with deeper, multi-step reasoning --- mirroring observations in FLOP counting.
Fang et~al.\ (2024) report similar deficits in systematic reasoning~\cite{fang2024large}.
This line of work motivates performance-centric reasoning about computational cost.

Other benchmarks touch performance indirectly.
\textsc{HumanEval} and \textsc{MBPP}~\cite{chen2021evaluating,austin2021program} occasionally reward efficiency (e.g., rejecting naive $O(n^2)$ code), but still emphasize correctness.
\textsc{LiveCodeBench}~\cite{jainLiveCodeBenchHolisticContamination2024} goes further by tracking runtime and memory usage to detect inefficiencies.
\textsc{gpuFLOPBench} complements these by targeting static performance reasoning --- quantifying cost without execution.

From the HPC and compiler perspective, ML has been used to model performance characteristics.
\textsc{Ithemal}~\cite{mendis2019ithemal} trained neural networks to predict CPU basic--block throughput, matching empirical timings and surpassing analytic models.
Similarly, \cite{arunavodeyRelativePerformancePrediction2024} fine-tunes an LLM to generate performance counter samples that are then passed to classical ML models (e.g., XGBoost, KNN, Linear Regression)  for execution time prediction.
Zixian et al. \cite{omniwiseBench} create a synthetic dataset of AMD HIP GPU kernels and fine-tune a Llama model to predict performance metrics, such as arithmetic intensity and cache hit rates.
Closer to our scope, Bolet et al.~\cite{bolet2025can} study whether LLMs can \emph{predict} parallel code performance in a single step, formulating a binary classification of kernels as compute--bound vs.\ memory--bound rather than regressing full performance metrics --- a complementary one--shot diagnostic distinct from our exact (or parametric) FLOP counting.
Together, these efforts suggest that while ML can capture useful performance signals, general-purpose LLMs have yet to internalize precise cost models; isolating static reasoning, as in \textsc{gpuFLOPBench}, makes this gap explicit.



\subsection{Machine Learning for Compiler Optimization}

Machine learning has long been applied to compiler optimization and performance modeling, offering useful context for our LLM–centric study.
Traditionally, compiler heuristics (e.g., inlining or loop–unrolling) were manually tuned; over the past two decades, researchers have replaced these heuristics with learned models that better explore complex optimization spaces.

One early milestone was \textsc{Milepost GCC}~\cite{fursin2011milepost}, which extracted static code features to predict good optimization flags.
Fursin~et~al.\ showed it could match or exceed \texttt{-O}~settings on \textsc{SPEC} benchmarks, demonstrating that ML can outperform fixed heuristics and paving the way for self–tuning compilers.
Later work extended this to predictive modeling of performance.
\textsc{DeepTune}~\cite{cummins2017deeptune}, for example, trained a deep network on raw OpenCL code to predict optimal device or thread–coarsening factors, outperforming hand–tuned baselines by 12–14\%.
Our approach is conceptually related but broader: we employ general–purpose LLMs for static FLOP–count estimation rather than task–specific tuning.

Learned cost models have also been successful in high–performance Domain Specific Languages (DSLs) such as \textsc{Halide} and \textsc{TVM}.
Adams~et~al.~\cite{adams2019halide} trained gradient–boosted trees to predict runtime in \textsc{Halide}'s auto–scheduler, achieving schedules up to twice as fast as prior methods and even surpassing expert–crafted ones.
Similar models in \textsc{TVM/Ansor} guide search over tensor–operator schedules.
These works prove that ML can capture performance characteristics difficult to encode manually.
However, today’s LLMs still struggle with much simpler static FLOP reasoning, suggesting the need for specialized fine–tuning or hybrid models integrating compiler–level cost modeling.

Graph–based learning offers another avenue.
\textsc{ProGraML}~\cite{cummins2021programl} represents LLVM~IR as graphs and uses GNNs for compiler–analysis tasks, showing that structured representations can improve reasoning over code.
Extending such graph–based representations into LLMs -- currently text–based -- remains an open challenge.

Overall, the ML–for–compilers community has achieved precise performance prediction through specialized models.
Our work bridges this domain with general LLMs, highlighting their current limitations and motivating hybrid approaches that combine LLM generality with learned compiler cost models.

\subsection{Existing Code Generation and Compilation Benchmarks}

A variety of benchmarks evaluate the code–generation, understanding, and compilation abilities of AI systems. 
We briefly position \textsc{gpuFLOPBench} within this landscape.

\textbf{General code–generation benchmarks.}
\textsc{HumanEval}~\cite{chen2021evaluating} and \textsc{MBPP}~\cite{austin2021program} evaluate functional correctness by requiring models to produce code that passes unit tests.
While these benchmarks advanced code synthesis research, they ignore efficiency --- a brute–force solution still receives full credit.
In contrast, our benchmark exposes such inefficiencies by quantifying FLOP cost.

\textbf{Diverse and adaptive benchmarks.}
\textsc{APPS}~\cite{hendrycks2021apps} and \textsc{CodeContests}~\cite{li2022alphacode} assess algorithmic reasoning via large sets of competitive-programming problems.
They implicitly reward efficiency through runtime limits but never inspect internal performance reasoning.
\textsc{gpuFLOPBench} differs by explicitly testing whether a model can infer computational cost statically.

\textbf{Holistic benchmarks.}
\textsc{LiveCodeBench}~\cite{jainLiveCodeBenchHolisticContamination2024} expands evaluation to tasks such as repair and explanation, dynamically revealing inefficiencies when code times out or exhausts memory.
Our benchmark complements this by isolating static performance reasoning --- asking models to predict cost directly, without execution.

\textbf{Code–compilation and optimization benchmarks.} 
Recent efforts like \textsc{KernelBench}~\cite{ouyang2025kernelbench}, \textsc{MultiKernelBench}~\cite{wen2025multikernelbench}, \textsc{Robust-kbench}~\cite{robustKBench}, and \textsc{TritonBench}~\cite{li2025tritonbench} test models on generating efficient CUDA, OpenCL, or Triton kernels.
Despite covering hundreds of ML operators, their results show that models are able to speed up codes, yet this is typically after an iterative refinement process as LLMs still struggle to consistently produce GPU kernels that compile and match proper input/output specs.
These studies probe generation and optimization together, whereas \textsc{gpuFLOPBench} isolates the reasoning component --- ``counting without running.''

Taken together, prior benchmarks emphasize correctness, compilation, or empirical speed, leaving algorithmic–cost reasoning largely untested.
\textsc{gpuFLOPBench} fills this gap by evaluating whether LLMs understand performance analytically.
In combination with functional and generative suites such as \textsc{HumanEval}, \textsc{APPS}, and \textsc{KernelBench}, it contributes to a comprehensive assessment of coding proficiency.
Ultimately, bridging correctness, optimization, and reasoning will be essential for assistants that not only write code but also critique and improve it.



\section{Our Benchmark: \benchName}
\label{sec:ourBenchmark}

We introduce \benchName as a new benchmark for evaluating the code analysis capabilities of LLMs.
It is designed to test the abilities of LLMs in predicting the FLOP counts of CUDA GPU kernels.
\benchName is a collection of 577 diverse real-world CUDA kernels sourced from the HeCBench suite \cite{hecbenchPaper, hecbenchGithub}.
For each kernel, \benchName maps \textbf{(1) Kernel Name}, \textbf{(2) Full Source Code}, \textbf{(3) Command line Input Arguments}, \textbf{(4) CUDA Launch Parameters}, and \textbf{(5) Compilation Commands} to empirically-profiled \textbf{single and double-precision FLOP counts} corresponding to the first execution of the kernel within its target executable.
This mapping essentially encompasses a static analysis task of FLOP count prediction for CUDA kernels.

We split the dataset into two categories, \textit{easy} and \textit{hard} subsets, with 204 and 373 kernels, respectively. 
This split is motivated by the types of kernels encountered within the HeCBench suite: those that can have their FLOP counts directly statically analyzed (\textit{easy} kernels), and those that, by their nature of not being statically analyzable, can incur hidden FLOP counts (hard kernels). 
We consider \textit{easy} codes as those which can be directly statically analyzed, in other words: codes that \textit{do not} require run-time or post-compilation information to reason about which operations a kernel will perform.
Given that for certain FLOP operations the compiler can add/remove FLOP instructions, or the entering of a particular branch (determined by some run-time value) could cause extra FLOP work to be done, these can make \textit{direct} static analysis of the CUDA C++ code challenging, as an analyzer must take into account post-compilation and dynamic behavior.
The presence of these attributes makes codes \textit{hard} to analyze, while their absence makes codes \textit{easy} to analyze.
A major contribution of this work is how we categorize each of the kernels in our dataset based on the various code attributes that make them \textit{hard}, allowing us to attribute common mispredictions to broad code features and make conclusions about where SoTA LLMs need improvement.

\noindentbf{CUDA Program Building.}
We built and automatically profiled 445 CUDA programs from the HeCBench suite, of which we were able to gather FLOP count data for 577 kernels within said programs.
Due to compilation or execution errors, we were unable to gather FLOP count data for some kernels.
For program compilation, we used LLVM clang 18.1.3 with CUDA version 12.6 and \texttt{-O3} optimization flag.
Compilation errors were typically caused by missing header files/dependencies, CUDA library dependency include order, or missing compiler flags.
Execution errors were caused by segmentation faults, out-of-memory errors, missing input data files, or some input failing to take a branch in the code that would launch a particular kernel.
We made best efforts to fix as many compilation and execution errors as possible, but ultimately decided we had a sizable number of kernels to move forward.

The target hardware used to gather FLOP count data was an NVIDIA RTX 3080 GPU with 10GB of RAM, where we compiled kernels for the Compute Capability (CC) \texttt{sm\_86} architecture.
Because compiling \textit{hard} codes for different architecture versions could lead to varying compiler CUDA SASS assembly output, we manually checked the compiler output across different compute capability versions (e.g., \texttt{sm\_52}, \texttt{sm\_86}, and \texttt{sm\_90}).
We found that the \textit{floating point operations were the same between versions}, where the only difference was a change in calling certain architecture-specific CUDA SASS instructions that did not change the FLOP counts.
From this finding, we could move forward using the FLOP counts gathered on our hardware, as they would be nearly identical across other CUDA GPUs. 

\newcommand{\treesitter}{\texttt{tree-sitter}\xspace}

\noindentbf{Kernel Profiling.} 
\autoref{fig:datasetCreation} depicts an overview of how we created our dataset.
We started with profiling as many kernels as we could from the HeCBench suite.
All our target executables are built without the \texttt{--use\_fast\_math} flag so as to get IEEE-compliant \cite{IEEEStandardFloatingPoint2019} single \textbf{(SP)} and double precision \textbf{(DP)} floating point operation calculations.
For each executable, HeCBench includes a \texttt{Makefile} with a \texttt{run} rule that contains the command line arguments to run said executable.
Using these command line defaults, we invoked each built executable using NVIDIA's Nsight Compute (\texttt{ncu}) profiler to gather SP and DP FLOP counts for the first invocation of each kernel.
We focus on the first invocation of each kernel as a simplification step, as some codes invoke CUDA kernels multiple times, once, or never.
This first-invocation simplification means that we expect the LLMs to be able to perform proper constant propagation of the command line arguments, all the way up through the first invocation of the desired kernel; any code post-kernel execution should be ignored by the LLMs.

\begin{figure}
    \centering
    \includegraphics[width=0.7\linewidth, trim={0 0.75cm 0 1.5cm}, clip]{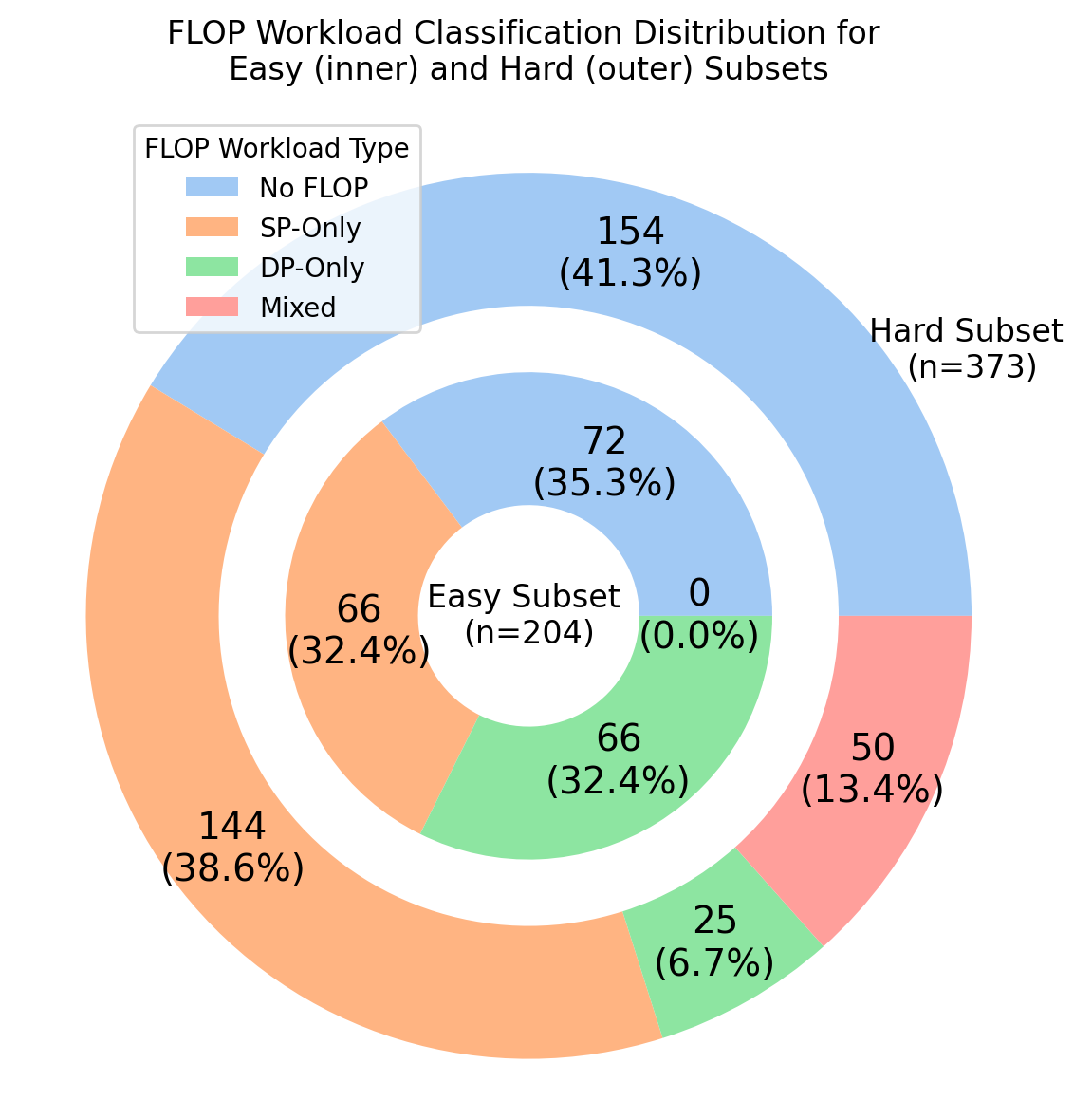}
    \caption{Distribution of FLOP workload type in \textit{easy} (inner ring) and \textit{hard} (outer ring) subsets of \benchName.
    The \textit{easy} subset lacks any mixed kernels that do both SP and DP FLOPs.
    The \textit{hard} subset is larger and has class imbalance due to the difficulties of balancing.
    }
    
    \label{fig:flopWorkloadDistrib}
    \vspace{0.1in}
\end{figure}

\noindentbf{Source Code Scraping.} 
We pair the profiled kernels with their corresponding source codes.
Initial attempts at automatically extracting kernels (and all their dependent code) were challenging due to the complexities of handling C++ templating, preprocessor defines, and data dependencies.
For simplicity, we chose to concatenate all the source files into one string for input.
This ensures that any code necessary for the proper analysis of each target kernel is included.
It also means that when querying LLMs about different kernels from the same source, we pass the same source code as input to the LLMs.
This places an additional expectation on the LLMs to correctly analyze the requested kernel in code that may contain multiple other kernels or irrelevant code.

\subsection{Source Code Feature Classification}
\label{sec:sourceCodeClassification}
From our expert knowledge, we knew that the CUDA kernels we chose to query could have \textit{hidden} FLOP counts that are only apparent post-compilation or at runtime.
We chose to categorize each of the scraped kernels based on the types of code \textit{attributes} they had that make it \textit{easy} or \textit{hard} to statically analyze FLOP counts.
These categorizations allow us to identify which code features challenge the LLMs the most, so we can conclude the current SoTA.
This code attribute annotation process was originally automated with LLMs, but after finding multiple errors, we converted the task to be done by a human using a web-based user interface (UI) for manual classification \cite{streamlitGithub}.
As seen in the top-middle portion of \autoref{fig:datasetCreation}, the UI would display the relevant source for a particular kernel, and the user could select between eight binary feature indicators corresponding to the code patterns of the kernel.

Five of these eight code attributes were recognized as common sources that complicate static analysis, and that, in turn, could introduce hidden FLOPs.
Below, we list these code attributes and discuss our expectations for LLMs tasked with analyzing these codes. 

\begin{itemize}
    \item Having \textbf{data-dependent branching/warp-divergence} where a conditional branch is dependent on a value read from memory or a value that could \textit{not} be calculated by constant propagation during static analysis. 
    This assumes the LLMs are powerful enough to know the distribution of values in memory that cause a branch to enter, thus being able to correctly calculate the number of CUDA threads that enter/pass a code region that may (or may not) contain FLOP operations.
    
    \item Using \textbf{intrinsic/special math functions} like \texttt{\_\_ddiv\_rd} or \texttt{cosf}, where the adherence to IEEE floating point precision standards causes the compiler to emit many implicit FLOP instructions (and data-dependent branches).
    This assumes the LLMs are powerful enough to know the number of new FLOPs the compiler has added, along with edge case value-handling branches it includes. 
    
    \item Performing \textbf{floating point division}.
    The CUDA compiler implements this operation in SASS using a reciprocal approximation coupled with iterative refinement, Fused-Multiply-Add (FMA) steps using Newton's method \cite{newtonRaphsonAlgorithmsFloatingpoint2010}.
    In particular, CUDA DP division includes an extra SP SASS instruction as a helper for efficient division implementation.
    With predicting FLOP counts of kernels using FLOP division, we assume that the LLMs are familiar with the operator implementation in the compiler and the FLOP data value distribution so as to properly estimate the number of iterative steps taken to compute a single DP division.
    
    \item Calling \textbf{external functions} from pre-compiled libraries (e.g., CUDA CCCL/CUB, \texttt{cuSparse}, \texttt{cuFFT}) easily introduces new FLOP operations, which places a burden on LLMs to be familiar with the various functions called and how many FLOPs they could introduce. 
    Given that libraries like \texttt{cuFFT} are often closed-source, this poses an additional challenge to LLMs in estimating FLOP counts without knowing exact implementation details.

    \item \textbf{Common subexpressions} are any expressions that are repeated throughout a kernel, which the compiler is designed to detect and ensure are computed only once to minimize the number of redundant operations.
    This means that LLMs need to be aware when duplicate FLOP work is being performed and count the work only once instead of multiple times.

\end{itemize}

The remaining three binary indicators were used to note other code features that our initial investigations found could be challenging for LLMs, but would not hinder direct static analysis of FLOP counts.
We describe the three code attributes below:

\begin{itemize}
    \item \textbf{Recursion} is used in a small set of CUDA programs.
    We expect that LLMs understand how deep a recursion goes and when a stopping condition is hit, especially when the recursion is performing FLOP operations.
    \item \textbf{\deviceKernel functions} are additional functions that can be called from a \globalKernel kernel.
    Programs can contain more than one of these, where the LLMs are expected to correctly identify caller-callee relationships and account for the potential FLOPs performed by these functions.
    \item \textbf{Warp divergence} is created on each branch instruction, where CUDA threads that do not enter the branch wait upon those that do.
    These warp divergence points are assumed to \textit{not} be data-dependent, whereby static analysis of their conditional statement is possible through adequate constant propagation of known kernel inputs.
    
\end{itemize}

\begin{table}[h]
\centering
\caption{Code attribute distribution for \easySubset and \hardSubset subsets}
\label{tab:attributeDistribution}
\resizebox{\columnwidth}{!}{%
\begin{tabular}{lrr}
\toprule
\textbf{Code Attributes} & \textbf{Easy Subset} & \textbf{Hard Subset} \\
\midrule
1: Has Warp Divergence & 181 (88.73\%) & 367 (98.39\%) \\
2: Has Data-Dependent Warp Divergence & 0 (0.00\%) & 191 (51.21\%) \\
3: Has FLOP Division & 0 (0.00\%) & 138 (37.00\%) \\
4: Calls external/library function & 0 (0.00\%) & 61 (16.35\%) \\
5: Calls \deviceKernel function & 38 (18.63\%) & 174 (46.65\%) \\
6: Calls intrinsic/special math function & 0 (0.00\%) & 231 (61.93\%) \\
7: Has Common Subexpression & 0 (0.00\%) & 33 (8.85\%) \\
8: Has Recursion & 0 (0.00\%) & 3 (0.80\%) \\
9: No Attributes & 23 (11.27\%) & 0 (0.00\%) \\
\bottomrule
\end{tabular}%
}
\end{table}

\begin{figure}
    \centering
    \includegraphics[width=\linewidth,trim={0 0 0 0.8cm},clip]{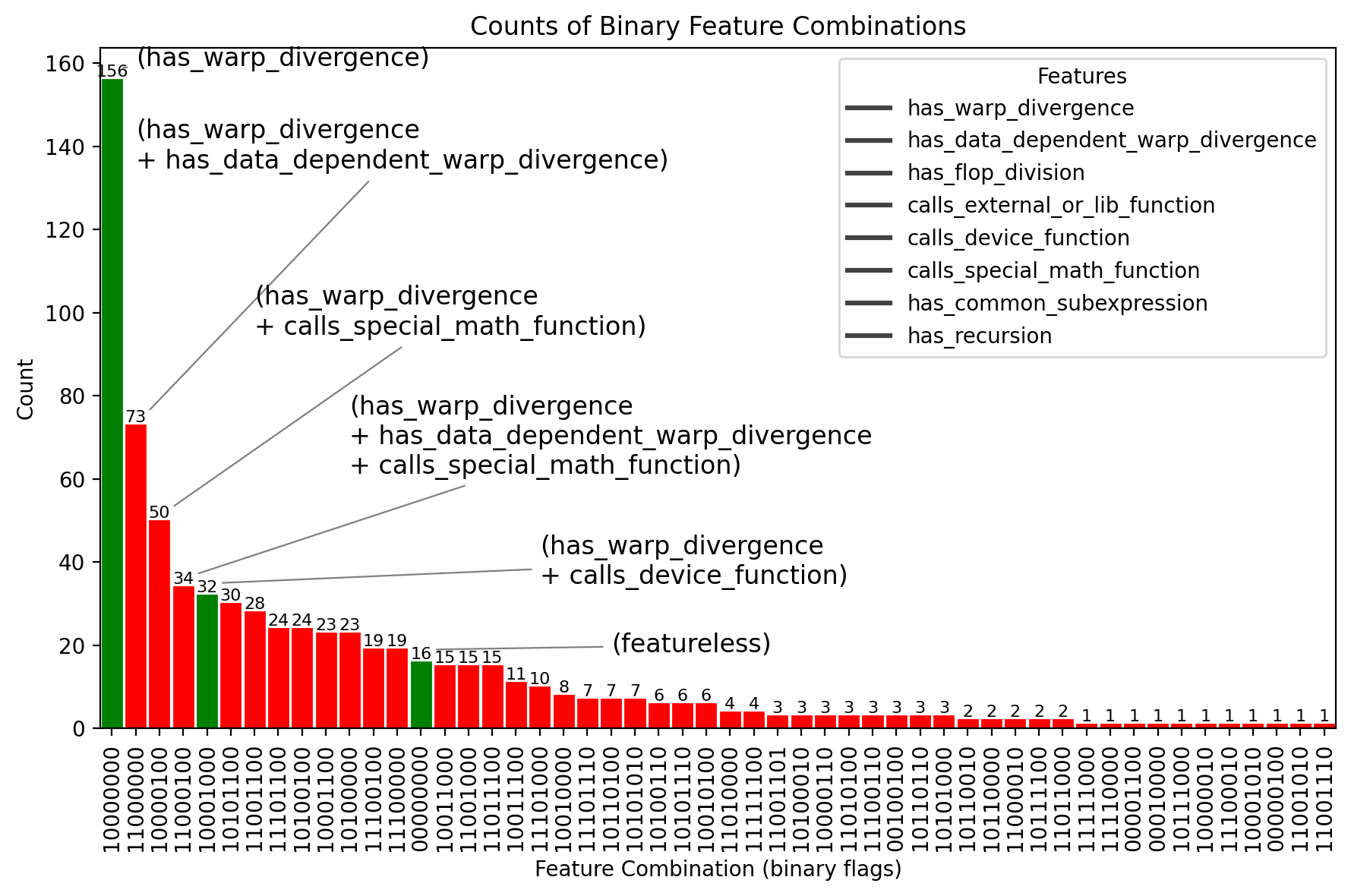}
    \caption{Counts of kernels with given combinations of execution attributes for 717 manually classified CUDA kernels from the HeCBench suite.
    Attribute combination indicators are represented as binary strings \textit{in the same order as the legend}.
    Green bars indicate attribute combinations that can be directly statically analyzed, while red bars are combinations that \textit{cannot} be directly statically analyzed and can thus engender indirect FLOPs.}
    \label{fig:code_feature_combinations}
\end{figure}

Given the above eight code features, we manually annotated 717 HeCBench kernels as having or not having them.
This categorization allowed us to create two subsets in our dataset; the \textit{easy} subset being kernels that contained \textit{none} of the five attributes that could introduce hidden FLOP counts; otherwise, kernels fell into the \textit{hard} subset.
\autoref{tab:attributeDistribution} shows the distribution of easy/hard features across the 577 kernels that made it into the final dataset. 
We note that the most common feature found in the sampled CUDA kernels is simple warp divergence, found in over 88\% of kernels in both subsets.
The attribute categories are not mutually exclusive; therefore, some kernels can have two or more attributes at the same time.

\subsection{Dataset Balancing \& Coalescing}

Although we had categorized all the CUDA kernels from the HeCBench executables we had built, we were unable to gather the FLOP counts for some kernels, due to execution errors mentioned earlier in this section. 
Of the kernels we had manually categorized \textit{and} profiled, we attempted to balance the \easySubset and \hardSubset subsets.
We balanced the \easySubset subset for the \spOnly and \dpOnly codes by simple string swapping of the \texttt{float} (SP) and \texttt{double} (DP) keywords in source code and compiler arguments, allowing us to use the same profiled FLOP counts without needing to re-build and profile the codes. 
This left us with 66 kernels in each set, as seen in \autoref{fig:flopWorkloadDistrib}.
The \hardSubset proved challenging to balance, as the \mixed codes would need to be re-profiled if we swapped datatypes, and we would need to drop kernels in the \spOnly and \dpOnly classes to match the 50 kernels of the \mixed class, which would severely shrink our dataset.
We opted to leave the \hardSubset subset as-is to simply reflect the distribution of FLOP workloads in scientific CUDA codes.
As a final dataset balancing step, we had tokenized all the scraped source codes and dropped outliers with more than 3e4 tokens to have similar distributions across FLOP workload categories while avoiding overwhelming the LLM input context windows.
\autoref{tab:nnz_flop_state_counts} shows the final distribution of the annotated attributes in each FLOP workload category of the dataset.

\begin{table*}[h]
\centering
\caption{Counts and percentages per code attribute across FLOP workload types in the \easySubset (\textbf{E}) and \hardSubset (\textbf{H}) subsets. Percentages are relative to the partition size. The \easySubset subset does not have \mixed FLOP workloads, hence the column is not shown.}
\label{tab:nnz_flop_state_counts}
\resizebox{\linewidth}{!}{%
\begin{tabular}{l|cc|c|cc|cc}
\toprule
\textbf{Code Attribute} & \textbf{DP-Only} (\textbf{E}) & \textbf{DP-Only} (\textbf{H}) & \textbf{Mixed} (\textbf{H}) & \textbf{SP-Only} (\textbf{E}) & \textbf{SP-Only} (\textbf{H}) & \textbf{No FLOPs} (\textbf{E}) & \textbf{No FLOPs} (\textbf{H}) \\
\midrule
1: Has Warp Divergence & 59 (89.39 \%) & 25 (100.00 \%) & 50 (100.00 \%) & 59 (89.39 \%) & 141 (97.92 \%) & 63 (87.50 \%) & 151 (98.05 \%) \\
2: Has Data-Dependent Warp Divergence & 0 (0.00 \%) & 14 (56.00 \%) & 27 (54.00 \%) & 0 (0.00 \%) & 66 (45.83 \%) & 0 (0.00 \%) & 84 (54.55 \%) \\
3: Has FLOP Division & 0 (0.00 \%) & 10 (40.00 \%) & 48 (96.00 \%) & 0 (0.00 \%) & 77 (53.47 \%) & 0 (0.00 \%) & 3 (1.95 \%) \\
4: Calls EXTERNAL or LIB function & 0 (0.00 \%) & 2 (8.00 \%) & 2 (4.00 \%) & 0 (0.00 \%) & 11 (7.64 \%) & 0 (0.00 \%) & 46 (29.87 \%) \\
5: Calls \texttt{\_\_device\_\_} function & 13 (19.70 \%) & 10 (40.00 \%) & 23 (46.00 \%) & 13 (19.70 \%) & 68 (47.22 \%) & 12 (16.67 \%) & 73 (47.40 \%) \\
6: Calls special math function & 0 (0.00 \%) & 14 (56.00 \%) & 19 (38.00 \%) & 0 (0.00 \%) & 105 (72.92 \%) & 0 (0.00 \%) & 93 (60.39 \%) \\
7: Has Common Subexpression & 0 (0.00 \%) & 2 (8.00 \%) & 16 (32.00 \%) & 0 (0.00 \%) & 14 (9.72 \%) & 0 (0.00 \%) & 1 (0.65 \%) \\
8: Has Recursion & 0 (0.00 \%) & 3 (12.00 \%) & 0 (0.00 \%) & 0 (0.00 \%) & 0 (0.00 \%) & 0 (0.00 \%) & 0 (0.00 \%) \\
9: No Attributes & 7 (10.61 \%) & 0 (0.00 \%) & 0 (0.00 \%) & 7 (10.61 \%) & 0 (0.00 \%) & 9 (12.50 \%) & 0 (0.00 \%) \\
\bottomrule
\end{tabular}
}
\end{table*}

\newcommand{\spflopexplanation}{\texttt{sp\_flop\_explanation}\xspace}
\newcommand{\spflopcount}{\texttt{sp\_flop\_count}\xspace}
\newcommand{\dpflopexplanation}{\texttt{dp\_flop\_explanation}\xspace}
\newcommand{\dpflopcount}{\texttt{dp\_flop\_count}\xspace}

\section{Experimental Setup}
\label{sec:experimental-setup}

\noindentbf{Sampled LLMs.}
\autoref{tab:sampledLLMs} contains a list of the LLMs we sampled for this work, including their capabilities and costs. 
We purposely target OpenAI's mini models because they are reasonably priced and allow us to get a sense of how LLMs have progressed over time.

\begin{table}[]
\centering
\caption{Information about sampled LLMs, their capabilities, and cost (in USD).
Sorted by release date.}
\label{tab:sampledLLMs}
\resizebox{0.9\columnwidth}{!}{%
\begin{tabular}{rccc}
\textbf{Model Name} &
  \textbf{Reasoning} &
  \textbf{\begin{tabular}[c]{@{}c@{}}Input/Output Cost\\ (\$/M tokens)\end{tabular}} &
  \textbf{\begin{tabular}[c]{@{}c@{}}Release\\ Date\end{tabular}} \\
gpt-4o-mini & & (0.15, 0.60) & 07-2024 \\
o3-mini & \checkmark  & (1.10, 4.40) & 01-2025 \\
gpt-5-mini & \checkmark & (0.25, 2.0) & 08-2025 \\
\end{tabular}%
}
\end{table}

\noindentbf{Prompt Design.}
Although there has been a surge of \textit{agentic} LLMs designed to autonomously analyze code and self-reflect in a feedback loop or agent pool, this approach is very costly, especially if restarts or errors are involved.
It is preferred that an LLM agent do as much as possible in a single query; thus, we focus on zero-shot prompting with minimal instructions for this work.

We query the previously mentioned LLMs using the prompt shown in \autoref{fig:mainPrompt}, with the full template and placeholder definitions cataloged in Appendix~\ref{sec:prompt-templates}.
The prompt was designed to inform the model of the prediction task, while supplying the minimum code information for prediction.
The information provided is: (1) kernel name, (2) command-line input arguments, (3) grid/block sizes, (4) compilation commands, (5) source code. 
Using the above 5 pieces of information, we expect the LLM to go through the process of constant propagation (using the command-line input arguments) up to the target kernel, followed by analysis of the inputs to the kernel and reasoning about the execution of the kernel's FLOP operations.
From some preliminary tests using the \textit{easy} dataset, we found some common mistakes made by the LLMs and thus included some explicit advice in the prompt to avoid said mistakes.
This same prompt is used to query both the \textit{easy} and \textit{hard} datasets.

This prompt is accompanied by an expected-output \textit{tool} call passed to the LLM, shown in \autoref{fig:toolCallPrompt} and likewise detailed in Appendix~\ref{sec:prompt-templates}.
As SoTA LLMs have been designed to make tool calls, we leverage the feature so as to have a consistent, structured output for analysis.
The expected output for the tool call consists of four fields:
\begin{itemize}
    \item \textbf{\texttt{sp/dp\_flop\_count}}.
    This field is the actual predicted FLOP count value; it is compared against the corresponding ground-truth profiled FLOP count.
    \item \textbf{\texttt{sp/dp\_flop\_explanation}}. 
    This field is filled with an explanation for how the LLM arrived at the reported \texttt{sp/dp\_flop\_count}.
    It serves as a way for us to inspect how a predicted \texttt{sp/dp\_flop\_count} was calculated.
\end{itemize}

\subsection{Evaluation Metrics}
We test each LLM's ability to accurately predict the SP and DP FLOP counts of 577 CUDA kernels.
For each kernel, we query each LLM 3 times repeatedly to capture any diversity in the LLM FLOP count predictions.
This prediction task can be broken down into two sub-tasks: \textbf{(1)} a FLOP workload classification task, and \textbf{(2)} a FLOP count prediction task.

\noindentbf{F1 Score \& MCC.} 
For evaluation task \textbf{(1)}, we quantify how well the LLMs can identify the workload that a kernel is doing as either: \noFLOPs, \spOnly, \dpOnly, or \mixed.
If a kernel has a \mixed workload (both SP \& DP FLOP counts are nonzero), but an LLM reports it as \dpOnly, this is considered a misprediction.
Similarly, if a kernel has an \spOnly workload, but an LLM reports it as \dpOnly.
This categorization is done by checking the zero (or non-zero) value of the JSON tool call LLM response values for the \texttt{sp\_flop\_count} and \texttt{dp\_flop\_count} fields.
Given that our dataset is imbalanced, we take a weighted-average F1 classification score, which accounts for the varying sizes (i.e., support) of each FLOP workload class.
This F1 score ranges between 0 (worst score) and 1 (best score).
As a summarization metric among all four FLOP workload types, we use Matthew's Correlation Coefficient (\textbf{MCC}) as it accounts for imbalanced datasets; an MCC score ranges from -1 to +1, with +1 indicating perfect predictions, and 0 being no better than random guessing.

\noindentbf{Mean Absolute Log Error (MALE).}
For evaluation task \textbf{(2)}, we quantify how close the FLOP count predictions are to the ground-truth empirical measurements from the NVIDIA profiler.
Given that the profiled FLOP values are within a wide range (from 0 to 1e12), we decided to put the errors on a log scale to better compare them.
We take the absolute value of the log-scaled ground-truth and predicted FLOP counts because under-predictions are negative, while over-predictions are positive, which could cause the average to incorrectly hang around an ideal error of 0.
We take the average of these errors as a summarization of all the predictions.
Due to how $\log_{10}(0)$ is undefined, we add 1 to all the FLOP counts, allowing our metric to capture the error of non-zero predictions on a zero-valued ground-truth FLOP count (and vice-versa).
The MALE formula is provided below:
\[
\text{MALE}(X) = \frac{1}{n} \sum_{i=1}^{n} \lvert \log_{10}(x_{i}+1) - \log_{10}(y_{i}+1) \rvert.
\]
Given a set $X$ of $n$ predictions ($x_i$) and ground-truth ($y_i$) FLOP count tuples $(x_i, y_i)$, the MALE error is the mean of the absolute differences between the two log-scaled values.
Intuitively, an error of 4.0 implies that the average prediction was 1e4 times off in either over- or under-prediction.
As an example, if the ground-truth FLOP count is 100 and the prediction was 1e6, the prediction is 1e4 times the ground-truth, and the MALE error would be 4.
Perfect predictions will have an error of 0.

\section{Results}
\label{section:results}

We pose the following research questions for the evaluation of LLMs in \benchName.
\begin{itemize}
    \item \textbf{RQ1}: How accurately can LLMs \emph{classify} the type of input code in terms of their FLOP workload?
    \item \textbf{RQ2}: How accurately can LLMs \emph{estimate} the FLOP counts of input kernels of varying workload types?
\end{itemize}

We answer \textbf{RQ1} in \autoref{sec:flopWorkloadClassificationResults} and \textbf{RQ2} in \autoref{sec:flow_count_prediction_results}.

\subsection{FLOP Workload Classification Results}
\label{sec:flopWorkloadClassificationResults}

\autoref{tab:mccScores} shows the F1 and MCC classification scores for each FLOP workload category (i.e., No FLOPs, SP-Only, DP-Only, and Mixed) across the \easySubset and \hardSubset data subsets. 
What we can notice is that for the kernels in the \easySubset subset, the reasoning models do fairly well in identifying when a code is (or isn't) doing FLOP work by the high scores in the \noFLOPs category.
Unfortunately, for \gptfouromini, it really struggles in identifying codes as not performing FLOPs or as performing DP FLOPs strictly.
From examining the model outputs, \gptfouromini easily mixes up the requested kernel for another or mistakenly assumes the datatypes of variables as floating point when they are actually integers or longs.
We see a massive improvement in correct workload classification with the reasoning models, with \gptfivemini scoring perfectly on the \easySubset subset.

\begin{table*}[!htb]
    \centering
    \caption{F1 and MCC classification scores for each class of the FLOP workload type predictions on both the \textit{easy} (\textbf{E}) and \textit{hard} (\textbf{H}) data subsets.
    For both F1 and MCC metrics, a value of 1 indicates perfect predictions.
    } 
    \label{tab:mccScores}
    \resizebox{\linewidth}{!}{%
    \begin{tabular}{rccc|cccc|cc}
\toprule
\textbf{Model Name} & \textbf{No FLOPs} (\textbf{E}) & \textbf{SP-Only} (\textbf{E}) & \textbf{DP-Only} (\textbf{E}) & \textbf{No FLOPs} (\textbf{H}) & \textbf{SP-Only} (\textbf{H}) & \textbf{DP-Only} (\textbf{H}) & \textbf{Mixed} (\textbf{H}) & \textbf{MCC} (\textbf{E}) & \textbf{MCC} (\textbf{H}) \\
\midrule
gpt-5-mini & 1.0000 & 1.0000 & 1.0000 & 0.9824 & 1.0000 & 0.9859 & 0.1507 & 1.0000 & 0.8104 \\
o3-mini & 0.9905 & 1.0000 & 1.0000 & 0.9862 & 0.9928 & 0.9933 & 0.1491 & 0.9902 & 0.8002 \\
gpt-4o-mini & 0.0897 & 0.9661 & 0.3431 & 0.1173 & 0.9940 & 0.2143 & 0.3784 & 0.2055 & 0.2381 \\
\bottomrule
\end{tabular}
    }
\end{table*}

Surprisingly, we see similar scores on the \hardSubset subset across all models.
This indicates that, regardless of code attributes that could cause indirect FLOPs, the LLMs have similar prediction behavior. 
From examining model outputs, we find that all three LLMs fail to recognize that kernels doing DP division incur implicit SP FLOPs, indicating a weakness in understanding how FLOPs are implemented on NVIDIA hardware.
It also highlights that LLMs take the classification task at face value: if they see explicit SP and DP FLOPs in a kernel, only then will they categorize it as \mixed, otherwise it is just another \spOnly or \dpOnly kernel.

Overall, the most recent reasoning models (\gptfivemini and \othreemini) perform very well on correctly categorizing \spOnly and \dpOnly FLOP workloads for both the \easySubset and \hardSubset subsets, except for miscategorizing \mixed codes due to their naive detection of SP and DP FLOPs and lack of understanding of how FLOP division is implemented on NVIDIA GPU hardware.
The final trend to note is that the MCC scores in the last two columns of \autoref{tab:mccScores} tell us at a high level that LLM capability in this task is improving over time, albeit slowly in 2025.
In 8 months, going from \gptfouromini to \othreemini showed a tremendous improvement in our F1 and MCC metrics, while in 7 months, going from \othreemini to \gptfivemini showed far less of a jump, yet still a gradual improvement.
This trend can most likely be attributed to the rise of reasoning models in early 2025, causing the performances of \othreemini and \gptfivemini to have such a boost compared to \gptfouromini.

\begin{figure*}
    \begin{subfigure}{\textwidth}
        \centering
        \includegraphics[width=\linewidth,trim={0 0.75cm 0 0},clip]{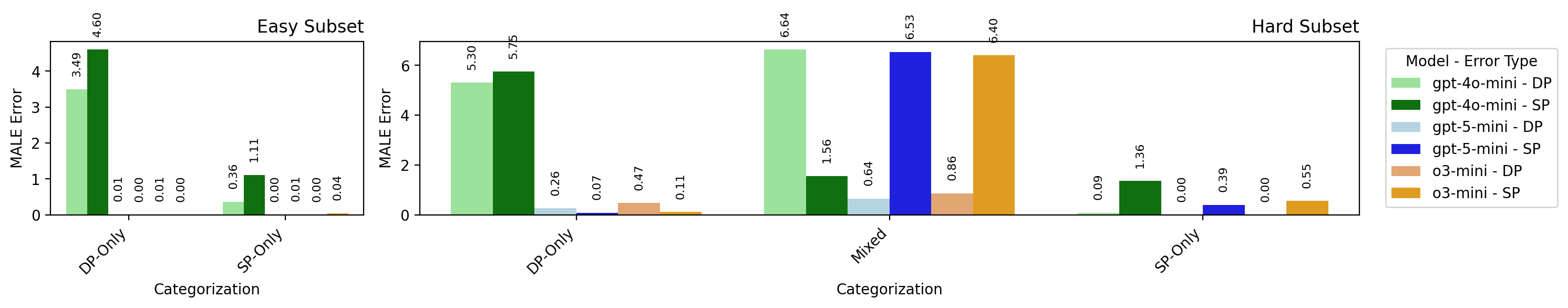}
        \caption{MALE error across FLOP workload categories (x-axis) on the \easySubset (left) and \hardSubset (right) subsets.
        The bars are colored according to each LLM sampled, with a darker color representing the error of SP predictions, and a lighter shade for DP prediction error.
        Explicit MALE error values are displayed above each bar.
        }
        \label{fig:maleScoresAcrossFLOPTypes}
    \end{subfigure} 
    \begin{subfigure}{\textwidth}
        \centering
        \includegraphics[width=\linewidth,trim={0 0.75cm 0 0},clip]{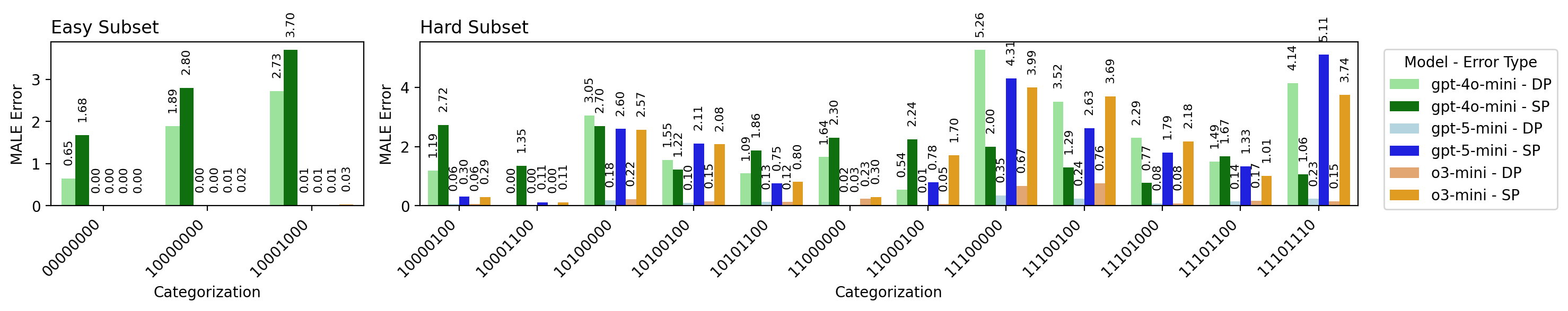}
        \caption{MALE error across code attributes on the \easySubset (left) and \hardSubset (right) subsets.
        Code attribute categorical combinations are shown as binary strings on the x-axis.
        }
        \label{fig:maleScoresAcrossCodeAttributes}
    \end{subfigure}
    \caption{Mean Absolute Log Error (MALE) results by subset on (a) FLOP workload categorization and (b) kernel code attributes. For MALE error, lower is better.}
    \label{fig:maleScores}
\end{figure*}

\subsection{FLOP Count Prediction Results}
\label{sec:flow_count_prediction_results}

For the task of FLOP count prediction, \autoref{fig:maleScores} shows the MALE errors across the \easySubset and \hardSubset subsets by FLOP workload categorization and code attributes.
We note that because the \noFLOPs class of kernels could cause very large errors due to a non-zero prediction being made on a zero-valued kernel, we purposely do not average in any results from the \noFLOPs category. 
Instead, it is better treated as a binary prediction task whose results are presented previously in \autoref{tab:mccScores}.

\noindentbf{Errors by FLOP Workload Type.}
From the dimension of \textit{FLOP workload types} in \autoref{fig:maleScoresAcrossFLOPTypes}, we can notice that across all the models, the prediction error scores are very close to 0 (the ideal) for \spOnly and \dpOnly codes predicted by the reasoning models on \textit{both} subsets.
It is when we consider the \mixed codes that the reasoning models have larger prediction errors, being off on average by a factor of 1e6.5 on predictions for SP-FLOPs, which can be attributed to how 96\% of the \mixed codes have floating point division operations that incur hidden SP FLOP counts as mentioned in \autoref{sec:sourceCodeClassification}.
The reasoning models simply return a prediction value of 0 for SP FLOPs of \mixed codes, causing the SP error to be so high.
Apart from the reasoning models, we can note that \gptfouromini has the largest errors, often mispredicting \dpOnly codes as \spOnly, causing the SP FLOP and DP FLOP errors to simultaneously be large for \dpOnly codes.
Lastly, for the \hardSubset subset, \gptfouromini has a strikingly low SP error on \mixed codes, mostly due to how it mistakenly reports DP FLOPs as SP FLOPs, giving the appearance that it is accounting for the hidden SP FLOPs from DP divisions, when in reality it often isn't counting DP FLOPs given the high DP error. 
Overall, the reasoning models score the best from the lens of FLOP workload types; however, they often fail to account for the SP FLOPs that come with DP FLOP division.

\noindentbf{Errors by Code Attributes.} 
Looking at the prediction errors from the perspective of the manually-annotated code attributes in both data subsets of \autoref{fig:maleScoresAcrossCodeAttributes}, we can immediately note that \gptfouromini has the largest prediction errors among all three LLMs.

For the \easySubset subset of kernels, the reasoning models have close to 0 error across the three different code attribute combinations in the subset for both SP and DP predictions.
\gptfouromini is able to have the least error on simple attribute-less codes (\texttt{00000000}), but its error increases as kernels introduce warp divergence (\texttt{10000000}) and call other \deviceKernel kernels (\texttt{10001000}).

The results in \autoref{fig:maleScoresAcrossCodeAttributes} on the \hardSubset subset of kernels show the MALE errors for the top 12 code attribute combinations.
We remind the reader that although a bar looks low, a MALE score of 0.3 indicates a 2x prediction error (e.g., if the ground-truth FLOP count is 1e4, the prediction was either 0.5e4 or 2.0e4).
All of the categories have the first bit present, indicating that all the kernels have warp divergence. 
There is a general trend that as the third bit (FLOP division) flips on, all three LLMs start to witness an increase in SP prediction error, with \gptfouromini having the largest errors for both SP and DP predictions.
This aligns with the fact that the DP FLOP division incurs many extra SP FLOPs that all the LLMs fail to account for.

With the second bit (data-dependent warp divergence) enabled, the case of \texttt{11000000} has the lowest reasoning-model errors around 0.02 for \gptfivemini and 0.27 for \othreemini.
As we enable any of the other bits, like bit six \texttt{11000100} (special math functions) or bit three \texttt{11100000} (FLOP division) and bit five \texttt{11101000} (calling device functions), the SP errors increase on the reasoning models.
When we have the most bits enabled in the case of \texttt{11101110}, the overall best model (\gptfivemini) has the highest SP error, mostly from having to predict on codes with common subexpressions (bit 7), special math functions (bit 6), and FLOP division (bit 3), all while dealing with data-dependent warp divergence (bit 2).

The addition of code attributes does not strictly increase the error for all cases. 
For example, going from \texttt{10000100} (warp divergence + special math functions) to \texttt{10001100} (adding \deviceKernel function calling) actually decreases the error for all models across SP and DP predictions.
The same decreasing trend is seen for the reasoning LLMs when adding \deviceKernel functions to \texttt{10100100} (warp divergence + division + special math functions) or \texttt{11100000}.
This is due to how breaking up code reuse into \textit{helper} functions makes a program more readable, which in turn helps the LLMs to perform more accurate FLOP calculations.

Lastly, there are decreases in errors of all the LLMs from \texttt{10100000} (warp divergence + division) to \texttt{10100100} (adding special math functions), which is counter-intuitive, as added calls to trig functions or intrinsics introduce more hidden FLOPs. 
This indicates additional variables that our categorization scheme is most likely not capturing.
We attribute this, for the most part, to the diversity of C++ CUDA codes, where our dataset may need to be augmented with more samples and categorizations (e.g., function templating) to capture a wider variety of codes and their attributes to better tease out prediction error differences.

\noindentbf{Takeaways.}
Although we cannot fully explain all the errors, two general trends remain clear: 
\textbf{(1)} The cost of implicit FLOPs (e.g., FLOP division, special math functions, common subexpressions) is \textit{not} being accounted for by SoTA LLMs, which causes them to largely fail the FLOP prediction task posed by our benchmark \benchName.
This highlights the bigger issue of LLMs not fully accounting for the complexity and workload of the code they emit.
\textbf{(2)} LLM capabilities have greatly improved over time with the advent of \textit{reasoning}, exemplified by the glaring differences in FLOP prediction errors between \gptfouromini and \othreemini; however, performance differences are marginally bigger going from \othreemini to \gptfivemini in about the same time span --- pointing to a need for the next big development in LLMs to have the same impact as reasoning did in early 2025.

\section{Limitations \& Discussion}
\label{sec:discussion}

\noindentbf{Hardware-specific ground truth.}
All ground-truth FLOP labels in \benchName were obtained on a single NVIDIA RTX~3080 GPU running CUDA~12.6.
While this setup offered relatively stable measurements for the first invocation of each kernel, it bakes NVIDIA-specific implementation details into the benchmark.
Division instructions that decompose into mixed-precision micro-ops, transcendental intrinsics expanded by the CUDA math libraries, or scheduler decisions around warp divergence all reflect this particular hardware and driver stack.
Consequently, our findings in \autoref{tab:mccScores} and \autoref{fig:maleScores} speak most directly to models' reasoning about NVIDIA-style execution.
Extending the benchmark to CPUs, AMD GPUs, or TPUs would require re-profiling and rethinking how vendor-specific optimizations manifest in the ground truth; short-lived kernels in those environments may also introduce considerably more measurement noise than we observed on the GPU.

\noindentbf{Dataset scope and annotations.}
\benchName currently covers 577 CUDA kernels drawn from HeCBench, which skews toward HPC-style workloads.
Many real systems interleave CUDA with host-side orchestration, vendor libraries, or domain-specific languages; none of those hybrid patterns are present.
Moreover, the ``easy''/``hard'' split hinges on manual annotation of eight execution attributes.
Although we double-checked the labels, subtle control-flow or template instantiations can slip through, and there are attributes (e.g., recursion) represented by only a handful of samples.
This limits the statistical power of the per-attribute MALE trends and leaves open whether the same failure modes would persist on computer-vision or ML kernels authored outside HeCBench.

\noindentbf{Evaluation protocol.}
Our current pipeline queries each model three times with the same zero-shot prompt and expects a structured tool-call response.
This choice isolates static reasoning ability, but it also underplays models or agents that rely on multi-turn self-reflection, auxiliary retrieval, or execution traces.

\noindentbf{Interpreting model failures.}
Although \benchName highlights where predictions go wrong, it does not yet reveal \emph{why}.
From the explanations returned by the models, we can only infer whether constant propagation stalled, loop bounds were miscomputed, or warp-level behavior was misunderstood.
A richer evaluation would capture intermediate reasoning artifacts -- symbol tables, unrolled loops, or explicit estimators of branch occupancy -- to distinguish arithmetic mistakes from control-flow mistakes.
Such instrumentation would also let us quantify how often implicit FLOPs (for example, DP division materializing SP operations) are recognized before the final count is emitted.

\noindentbf{Implications for future systems.}
Despite these limitations, the consistent gains from \gptfouromini to \othreemini and \gptfivemini suggest that reasoning-focused training helps LLMs track FLOP workloads more faithfully.
At the same time, the universal struggle with \mixed kernels underlines that models lack a mental model of hardware microcode: they see explicit SP and DP arithmetic, but miss the implicit conversions inserted by the compiler.
This gap points toward hybrid static–dynamic approaches where an LLM augments symbolic reasoning with lightweight profiling or reference to ISA-level templates.
Coupling \benchName with such tooling -- and widening it beyond NVIDIA GPUs -- would turn the benchmark from a point-in-time snapshot into a scaffold for building truly performance-aware coding assistants.


\section{Acknowledgments}
This research was supported by Code Metal Inc. and the Pazy Foundation.
This work was performed under the auspices of the U.S. Department of Energy by Lawrence Livermore National Laboratory under Contract DE-AC52-07NA27344 (LLNL-CONF-2013245).
The views and opinions of the authors do not necessarily
reflect those of the U.S. government or Lawrence Livermore National Security, LLC neither of whom nor any of
their employees make any endorsements, express or implied
warranties or representations or assume any legal liability
or responsibility for the accuracy, completeness, or usefulness of the information contained herein. 
The United States Government retains, and
the publisher, by accepting the article for publication, acknowledges that the United States Government
retains a non-exclusive, paid-up, irrevocable, world-wide license to publish or reproduce the published form
of this manuscript, or allow others to do so, for United States Government purposes.

\bibliography{references}

\newpage
\appendix

\clearpage
\section{Appendix: Prompt Templates}\label{sec:prompt-templates}

We include the exact instruction prompt and structured response schema used for every LangGraph query to ensure our results are reproducible. The zero-shot template in \autoref{fig:mainPrompt} foregrounds the information the model must reason over while keeping the request focused on FLOP counting. The highlighted placeholders denote per-kernel values supplied programmatically, allowing us to reuse a single prompt across both the \easySubset{} and \hardSubset{} partitions with consistent guidance.

\noindent\textbf{Instruction prompt.} The system message primes the model as a FLOP-counting assistant and calls out frequent failure modes we observed during pilot runs, such as overlooking unary negations or misinterpreting command-line constants. The accompanying human message expands the context with the kernel metadata needed for static reasoning: launch configuration, build flags, and concatenated source files. Together, these fields encourage the model to propagate inputs to the target function before enumerating floating-point ops.

\noindent\textbf{Structured response schema.} To simplify downstream analysis, we require the model to emit a tool call matching the Pydantic schema in \autoref{fig:toolCallPrompt}. Separating the count fields from their natural-language explanations helps us compare predictions against ground truth while still retaining the model's rationale for error analysis.

\begin{figure}[h!]
    \centering
    \begin{tcolorbox}[
        colframe=OliveGreen!70!black,     
        colback=white,                    
        coltitle=white,                   
        colbacktitle=OliveGreen!80!black,
        title=Expected LLM Output: LangGraph Tool Call,
        fonttitle=\bfseries,
        enhanced,
        sharp corners=south,
        boxrule=0.4mm,
        width=1.0\linewidth,
        before upper={\parindent0pt},    
        fontupper=\small,
        fontlower=\tiny,
        left=1pt,
        right=1pt,
        top=1pt,
        bottom=0pt,
        arc=2mm,
        drop shadow=black!50!white,
    ]

\begin{minted}[breaklines, fontsize=\tiny]{python}
class FLOPCounts(BaseModel):
    sp_flop_explanation: str = Field(..., description="Explanation of how the single-precision floating point operations (SP-FLOP) count was calculated. This should include the reasoning behind the number of operations performed in the kernel, including any relevant loop iterations and warp divergence region executions.")

    sp_flop_count: int = Field(..., description="Total number of single-precision floating point operations (SP-FLOP) performed by the kernel. Accounting for the number of threads, loop iterations, and warp divergence region executions.")

    dp_flop_explanation: str = Field(..., description="Explanation of how the double-precision floating point operations (DP-FLOP) count was calculated. This should include the reasoning behind the number of operations performed in the kernel, including any relevant loop iterations and warp divergence region executions.")

    dp_flop_count: int = Field(..., description="Total number of double-precision floating point operations (DP-FLOP) performed by the kernel. Accounting for the number of threads, loop iterations, and warp divergence region executions.")
\end{minted}

    \end{tcolorbox}
    \caption{LangGraph tool call structure passed to an LLM upon invocation. 
    The field descriptions and datatypes help the LLM to accurately complete the tool call.
    }
    \label{fig:toolCallPrompt}
\end{figure}

\begin{figure}[h!]
    \centering
    \begin{tcolorbox}[
        colframe=OliveGreen!70!black,     
        colback=white,                    
        coltitle=white,                   
        colbacktitle=OliveGreen!80!black,
        title=LLM Zero-shot Prompt,
        fonttitle=\bfseries,
        enhanced,
        sharp corners=south,
        boxrule=0.4mm,
        width=1.0\linewidth,
        before upper={\parindent0pt},    
        fontupper=\small,
        fontlower=\tiny,
        left=1pt,
        right=1pt,
        top=1pt,
        bottom=0pt,
        arc=2mm,
        drop shadow=black!50!white,
    ]

\centerline{\texttt{[SYSTEM MESSAGE]}}

You are an expert CUDA source code FLOP counting assistant. For a given target CUDA kernel, you will be given: 

\begin{itemize}[label={}]
    \item A) The Target Kernel Name
    \item B) Commandline Input Arguments
    \item C) Grid and Block Size Launch parameters. 
    \item D) Source Compilation Commands (including any relevant preprocessor defines)
    \item E) Concatenated Source Code Files
\end{itemize}

Your task is to analyze the code and accurately determine the number of single-precision (SP-FLOP) and double-precision (DP-FLOP) floating point operations (FLOP) performed by the kernel during its FIRST execution invocation.

When counting FLOPs, be sure to remember the following:

\begin{itemize}[label={}]
    \item  - unary negation of a float/double (e.g., -x) DOES count as a floating point operation.
    \item - code comments may incorrectly state the number of FLOPs, do not trust them, instead calculate them yourself
    \item - other floating point datatypes like FP16 (half2) or FP8 should not be counted as SP or DP FLOPs
    \item - commandline input arguments may not be used directly in the kernel function call, they may be passed through other functions or used to compute other values
    \item - if the target kernel is templated, be sure to only report on the execution of its FIRST instantiation
\end{itemize}

Provide a detailed explanation of how you arrived at the SP-FLOP and DP-FLOP counts, including any assumptions or simplifications you made during your analysis. Report the final SP-FLOP and DP-FLOP counts using the \spflopcount, \dpflopcount, \spflopexplanation and \dpflopexplanation fields in your response.

\centerline{\texttt{[HUMAN MESSAGE]}}

\begin{itemize}
    \item Target Kernel Name: \hl{\{kernel\_name\}}
    \item Commandline Input Arguments: \hl{\{exec\_args\}}
    \item Grid Size: \hl{\{grid\_size\}}
    \item Block Size: \hl{\{block\_size\}}
    \item Total Number of Threads: \hl{\{total\_num\_threads\}}
    \item Compilation Commands: \\
         \hl{\{compile\_commands\}}
\end{itemize}

Please return the completed FLOP counts and explanations in the following fields: \spflopcount, \dpflopcount, \spflopexplanation, \dpflopexplanation. 
Source code: \\
\hl{\{source\_code\}}

    \end{tcolorbox}
    \caption{Prompt used to make LLM queries. 
    It is composed of two messages, a \textit{system message} with general instructions, and a \textit{human} message containing all the details used for prediction.
    Highlighted text indicates values changed between invocations.
    We used the LangGraph \cite{langgraphGithub} library for prompting.
    }
    \label{fig:mainPrompt}
\end{figure}

%

\end{document}